\documentclass[letterpaper,12pt]{article}

\usepackage[numbers]{natbib}
\usepackage{bibentry}
\usepackage{graphics,graphicx,psfrag,color,float, hyperref}
\usepackage{epsfig,amsfonts,amssymb,amsthm,url}
\usepackage{amsmath}
\usepackage{amssymb}
\usepackage{graphicx,bibentry}
\newtheorem{theorem}{Theorem}
\newtheorem{lemma}{Lemma}

\newcommand{\myexp}{{\mathrm{e}}}
\providecommand{\abs}[1]{\lvert#1\rvert}
\newcommand{\beq}{\begin{equation}}
\newcommand{\enq}{\end{equation}}
\newcommand{\beqa}{\begin{eqnarray}}
\newcommand{\enqa}{\end{eqnarray}}
\newcommand{\beqan}{\begin{eqnarray*}}
\newcommand{\enqan}{\end{eqnarray*}}
\newcommand{\cA}{{\cal A}}
\newcommand{\ignore}[1]{}
\newcommand{\nobibentry}[1]{{\let\nocite\ignore\bibentry{#1}}}

\begin {document}

\title{Entropy power inequality for a family of discrete random variables }

\author{
{\bf Naresh Sharma}  ~~ {\bf Smarajit Das} ~~  {\bf Siddharth Muthukrishnan} \\
Tata Institute of Fundamental Research \\
Mumbai 400 005, India \\
Email: \texttt{nsharma@tifr.res.in}
}

\date{\today}

\maketitle

\maketitle
\begin{abstract}
It is known that the Entropy Power Inequality (EPI) always holds if the random variables have density.
Not much work has been done to identify discrete distributions for which the inequality holds with
the differential entropy replaced by the discrete entropy.
Harremo\"{e}s   and Vignat showed that it holds for the pair
$(B(m,p), B(n,p))$, $m,n \in \mathbb{N}$,
(where $B(n,p)$ is a Binomial distribution with $n$ trials each with success probability
$p$) for $p = 0.5$ \nobibentry{Harrem}.
In this paper, we considerably expand the set of Binomial distributions for which the inequality holds 
and, in particular, identify $n_0(p)$ such that for all $m,n \geq n_0(p)$, the
EPI holds for $(B(m,p), B(n,p))$. We further show that the EPI holds for the discrete random variables
that can be expressed as the sum of $n$ independent identical distributed (IID) discrete
random variables for large $n$.
\end{abstract}

\section{Introduction}
The Entropy Power Inequality  
\begin{equation}
\myexp^{2h(X+Y)} \geq \myexp^{2h(X)}+\myexp^{2h(Y)}
\end{equation}
holds for independent random variables $X$ and $Y$ with densities, where $h(\cdot)$ is the
differential entropy. It was first stated by Shannon
in Ref. \cite{shannon},   and the proof was given by Stam and Blachman \cite{blachman}.
See also Refs. \cite{costas,dembo,johnson,arstein,tulino,verdu-guo,mokshay}.

This inequality is, in general, not true for discrete distributions where the differential entropy is
replaced by the discrete entropy. For some special cases (binary random variables with modulo $2$
addition), results have been provided by Shamai and Wyner in Ref. \cite{shlomo}.

More recently,
Harremo\"{e}s   and Vignat  have shown that this inequality will hold if $X$ and $Y$ are $B(n,1/2)$ and
$B(m,1/2)$ respectively for all $m,n$ \cite{Harrem}. Significantly, the convolution operation to get
the distribution of $X+Y$ is performed over the usual addition over reals and not over finite fields.

Recently, another approach has been expounded by Harremo\"{e}s   et. al. \cite{harr} and by
Johnson and Yu \cite{yu}, wherein they interpret
R{\'{e}}nyi's thinning operation on a discrete random variable as a
discrete analog of the scaling operation for continuous random variables.  They
provide inequalities for the convolutions of thinned discrete random variables that can be
interpreted as the discrete analogs of the ones for the continuous case.

In this paper, we take a re-look at the Harremo\"{e}s   and Vignat \cite{Harrem} result for the
Binomial family and extend it for all $p \in (0,1)$. We show that there always exists   an   $n_0(p)$
that is a function of $p$, such that for all $m,n \geq n_0(p)$,  
\begin{equation}
\myexp^{2H[B(m+n,p)]}\geq \myexp^{2H[B(m,p)]}+\myexp^{2H[B(n,p)]},
\end{equation}
where $H(\cdot)$ is the discrete entropy.
The result in Ref. \cite{Harrem} is a special case of our result since we obtain $n_0(0.5) = 7$
and it can be
checked numerically by using a sufficient condition that the inequality holds for $1\leq m,n\leq 6$.

We then extend our results for the family of discrete random variables that
can be written as the sum of $n$ IID random variables
and show that for large $n$, EPI holds.

We also look at the semi-asymptotic case for the distributions $B(m,p)$ with $m$ small and
$B(n,p)$ with $n$ large. We show that even when $n$ is large, there may exist some $m$ such
that EPI may not hold.

Lastly, we show that how the EPI for the discrete case can be interpreted as an
improvement to the bounds given by Tulino and Verd\'{u} for special cases \cite{tulino}.

\section{EPI for the Binomial distribution}

Our aim is to have an estimate on the threshold $n_0(p)$ such that  
\begin{equation}
\myexp^{2H[B(m+n,p)]}\geq \myexp^{2H[B(m,p)]}+\myexp^{2H[B(n,p)]},
\end{equation}
holds for all $m,n\geq n_0(p)$.

It is observed that $n_0(p)$ depends on the skewness of the associated Bernoulli distribution. Skewness of a probability distribution is defined as  $\kappa_3/\sqrt{\kappa_2^3}$
where  $\kappa_2$ and $\kappa_3$   are respectively the   second and third cumulants
of the   Bernoulli distribution $B(1,p)$, and it turns out to be $(2p-1)/\sqrt{p(1-p)}$. Let
\begin{equation}
\omega(p)=  \frac{{(2p-1)}^2}{p(1-p)}.
\end{equation}
We find an expression for $n_0(p)$ that depends on $\omega(p)$.
The following theorem, known as Taylor's theorem, will be useful   for this purpose (see for
example p. 110 in Ref. \cite{rudin}).
 \begin{theorem}[Taylor]
\label{taylor}
Suppose $f$ is a real function on $[a,b]$, $n$ $\in$ $\mathbb{N}$, the
$(n-1)$th derivative of $f$ denoted by $f^{(n-1)}$ is continuous on $[a,b]$, and
$f^{(n)}(t)$ exists for all $t$ $\in$ $(a,b)$. Let $\alpha$, $\beta$ be distinct points
of $[a,b]$, then there exists a point $y$ between $\alpha$ and $\beta$ such that
\begin{equation}
f(\beta)=f(\alpha)+\sum_{k=1}^{n-1}\frac{f^{(k)}(\alpha)}{k!}{(\beta-\alpha)}^k+
\frac{f^{(n)}(y)}{n!}{(\beta-\alpha)}^n.
\end{equation}
\end{theorem}
For $0\leq p\leq 1$, let $H(p)$ denote the discrete entropy of a Bernoulli distribution with probability of success $p$, that is, $H(p)\triangleq -p\log (p) - (1-p) \log(1-p)$. We shall use the natural
logarithm throughout this paper. Note that we earlier defined $H(\cdot)$
to be the discrete entropy of a discrete random variable. The definition to be used would be
amply clear from the context in what follows. Let
\begin{eqnarray}
\label{hhat}
\hat{H}(x)\triangleq H(p)-H(x), ~~~ x\in (0,1).
\end{eqnarray}
Note that $\hat{H}(x)$ satisfies the assumptions in the Theorem \ref{taylor} in $x \in (0,1)$. Therefore,
we can write  
\begin{equation}
\hat{H}(x)=\hat{H}(p)+\sum_{k=1}^{n-1}\frac{\hat{H}^{(k)}(p)}{k!}{(x-p)}^k+\frac{\hat{H}^{(n)}(x_1)}{n!}{(x-p)}^n,
\end{equation}
for some $x_1\in (x,p)$.
Note that $\hat{H}(p)=0$ and
\begin{eqnarray}
\label{coeff}
 F^{(k)}(x) \triangleq  \frac{\hat{H}^{(k)}(x)}{k!} =
\begin{cases}
 \log (x)-\log(1-x), & \mbox{ if } k=1, \\
 \frac{1}{k(k-1)}\left[{(1-x)}^{-(k-1)} +{(-1)}^{k} {x}^{-(k-1)}\right], & \mbox{ if } k\geq 2. 
 \end{cases}
\end{eqnarray}
For even $k$, $F^{(k)}(x)\geq 0$ for all $x\in (0,1)$, and hence, 
\begin{eqnarray}
\label{bound}
\hat{H}(x)\geq\sum_{k=1}^{2l+1}F^{(k)}(p){(x-p)}^k
\end{eqnarray}
for all $x\in (0,1)$ and any non-negative integer $l$.
The following useful identity would be employed at times
\begin{eqnarray}
\label{bin1}
\log (2)-H(p)=\sum_{\nu=1}^{\infty}\frac{2^{2\nu}}{2\nu(2\nu-1)}{\left(p-\frac{1}{2}\right)}^{2\nu}.
\end{eqnarray}

Let $P\triangleq \{p_i\}$ and $Q\triangleq \{q_i\}$ be two probability measures over a finite alphabet
$\cA$. Let $C^{(p)}(P,Q)$ and $\triangle_{\nu}^{(p)}(P,Q)$ be  measures  of discrimination defined as
\begin{eqnarray}
C^{(p)}(P,Q)&\triangleq&pD(P\parallel M)+qD(Q\parallel M),\\
\triangle_{\nu}^{(p)}(P,Q)&\triangleq&\sum_{i\in A}\frac{\abs{pp_i-qq_i}^{2\nu}}{(pp_i+qq_i)^{2\nu-1}},
\end{eqnarray}
where
\beqan
M & \triangleq & pP+qQ, \\
q & \triangleq & 1-p.
\enqan
These quantities are generalized capacitory discrimination and triangular discrimination of
order $\nu$ respectively that were introduced by Tops\o e \cite{topsoe}.

The following theorem relates $C^{(p)}(P,Q)$ with $\triangle_{\nu}^{(p)}(P,Q)$
and would be used later to derive an expression for $n_0(p)$. It generalizes Theorem 1
in Ref. \cite{topsoe}.
\begin{theorem}
\label{capa}
Let $P$ and $Q$ be two distributions over the alphabet $\cA$ and $0 < p < 1$. Then  
\begin{eqnarray}
C^{(p)}(P,Q)=\sum_{\nu=1}^{\infty}\frac{\triangle_{\nu}^{(p)}(P,Q)}{2\nu(2\nu-1)}-[\log (2)-H(p)].
\end{eqnarray}
\end{theorem}
\begin{proof}
Let
\begin{equation}
m_i=pp_i+qq_i,~~~\epsilon_i=\abs{2pp_i-m_i},~~~k_i=\frac{m_i}{\epsilon_i}.
\end{equation}
We have
\beq
\frac{1}{k_i}=\frac{\abs{pp_i-qq_i}}{pp_i+qq_i}
\enq
and $0 \leq 1/k_i \leq 1$. We have
\begin{eqnarray}
C^{(p)}(P,Q)&=&p\sum_{i\in \cA}p_i\log \left( \frac{p_i}{m_i} \right) +
q\sum_{i\in \cA}q_i \log \left( \frac{q_i}{m_i} \right) \\
& = & p\sum_{i\in \cA}p_i\log \left( \frac{pp_i}{m_i} \right) +
q\sum_{i\in \cA}q_i \log \left( \frac{qq_i}{m_i} \right) +H(p) \\
&\stackrel{a}{=}&\sum_{i\in \cA}\frac{m_i+\epsilon_i}{2}\log \left( \frac{m_i+\epsilon_i}{m_i} \right) +
\sum_{i \in \cA} \frac{m_i-\epsilon_i}{2}\log \left( \frac{m_i-\epsilon_i}{m_i} \right) \nonumber\\
&&	~~~~ -[\log (2)-H(p)] \\
&=&\sum_{i\in \cA}\frac{1}{2}\epsilon_i(1+k_i)\log \left(1+\frac{1}{k_i}\right)+\frac{1}{2}\epsilon_i
(k_i-1)\log \left(1-\frac{1}{k_i}\right)\nonumber\\
&& ~~~~ -[\log (2)-H(p)] \\
&=&\sum_{i\in A}\epsilon_ik_i\left[\log(2)-H\left(\frac{1}{2}+\frac{1}{2k_i}\right)\right]-[\log(2)-H(p)]\\
&\stackrel{b}{=}
&\sum_{i\in \cA}\epsilon_i\sum_{\nu=1}^{\infty}\frac{1}{2\nu(2\nu-1) k_i^{2\nu-1}} -[\log (2)-H(p)]\\
&=&\sum_{\nu=1}^{\infty}\frac{\triangle_{\nu}^{(p)}(P,Q)}{2\nu(2\nu-1)}-[\log (2)-H(p)],
\end{eqnarray}
where $a$ follows by taking two cases $2pp_i > m_i$ and $2pp_i \leq m_i$, and $b$ follows
from \eqref{bin1}.
\end{proof}

Let $X^{(n)}$ be a discrete random variable that can be written as
\beq
X^{(n)} = Z_1+Z_2+\cdots+Z_n,
\enq
where $Z_i$'s are IID random variables.
We note that when   $X^{(n)}$ is defined as above, we have $X^{(n)} + X^{(m)} =X^{(n+m)}$.
Let $Y_n \triangleq \myexp^{2[H(X^{(n)})]}$.  
We first use a lemma due to Harremo\"{e}s   and Vignat \cite{Harrem}.

\begin{lemma}[Harremo\"{e}s   and Vignat \cite{Harrem}]
\label{superadd}
If $Y_n/n$ is increasing, then $Y_n$ is super-additive, i.e., $Y_{m+n} \geq Y_m+Y_n$.
\end{lemma}

It is not difficult to show that this is a sufficient condition for the EPI to hold \cite{Harrem}.
By the above lemma, the inequality 
\begin{equation}
\label{pos}
H(X^{(n+1)})-H(X^{(n)}) \geq \frac{1}{2} \log \left( \frac{n+1}{n} \right)
\end{equation}
is sufficient for EPI to hold.

Let $X^{(n)} = B(n,p)$. We have
\begin{eqnarray}
P_{X^{(n+1)}}(k+1) = p P_{X^{(n)}}(k) + q P_{X^{(n)}}(k+1).
\end{eqnarray}
Define a random variable ${X^{(n)}}+1$ as
\begin{eqnarray}
P_{{X^{(n)}}+1}(k+1)=P_{X^{(n)}}(k).
\end{eqnarray}
for all $k \in \{0,1,\cdots,n\}$. Hence,  using $H(X^{(n)}+1) = H(X^{(n)})$, we have
\begin{eqnarray}
P_{X^{(n+1)}}&=&pP_{X^{(n)}+1}+ q P_{X^{(n)}},\\
 H(X^{(n+1)})&=&pH(X^{(n)}+1)+ q H(X^{(n)})+pD(P_{X^{(n)}+1}\parallel P_{X^{(n+1)}})
 + q D(P_{X^{(n)}}\parallel P_{X^{(n+1)}}) ~~~~~~  \\
&=&H(X^{(n)})+pD(P_{X^{(n)}+1}\parallel P_{X^{(n+1)}})+ q D(P_{X^{(n)}}\parallel P_{X^{(n+1)}}).
\end{eqnarray}
Therefore,
\beq
\label{dummy2}
H(X^{(n+1)})=H(X^{(n)})+C^{(p)}(P_{X^{(n)}+1},P_{X^{(n)}}).
\enq
We now derive the lower bound for $C^{(p)}(P_{X^{(n)}+1},P_{X^{(n)}})$.

\begin{lemma}
\label{JSD}
For $l \in \mathbb{N}$,
\begin{eqnarray}
C^{(p)}(P_{X^{(n)}+1},P_{X^{(n)}}) & = & \sum_{i=0}^{n+1}\hat{H} \left(\frac{i}{n+1} \right)
P_{X^{(n+1)}}(i), \\
C^{(p)}(P_{X^{(n)}},P_{X^{(n)}+1}) & \geq & \sum_{k=1}^{2l+1}F^{(k)}(p)(n+1)^{-k} \mu_k^{(n+1)},
\end{eqnarray}
where $\mu_k^{(n)}$ is the $k$-th central moment of $B(n,p)$, i.e.,
\begin{equation}
\mu_k^{(n)}=\sum_{i=0}^{n}{(i-np)}^kP_{X^{(n)}}(i).
\end{equation}
\end{lemma}
\begin{proof}
Let $P=X^{(n)}+1$ and $Q=X^{(n)}$. We have  
\begin{eqnarray}
\frac{\abs{pp_i-qq_i}}{pp_i+qq_i}&=&
\frac{p\binom{n}{i-1} p^{i-1}q^{n-i+1}-q\binom{n}{i} p^{i}q^{n-i}}
{p\binom{n}{i-1}p^{i-1}q^{n-i+1}+q\binom{n}{i}p^{i}q^{n-i}}\\
&=&\frac{ \left[\binom{n}{i-1}-\binom{n}{i} \right]p^{i}q^{n-i+1}}
{ \left[\binom{n}{i-1}+\binom{n}{i} \right]p^{i}q^{n-i+1}}\\
&=&\frac{2i-n-1}{n+1},
\end{eqnarray}
\begin{eqnarray}
\triangle_{\nu}^{(p)}(P_{X^{(n)}+1},P_{X^{(n)}})&=&
\sum_{i=0}^{n+1}\left(\frac{2i-n-1}{n+1}\right)^{2\nu}P_{X^{(n+1)}}(i)\\
&=&\left(\frac{2}{n+1}\right)^{2\nu}\sum_{i=0}^{n+1}\left(i-\frac{n+1}{2}\right)^{2\nu}P_{X^{(n+1)}}(i).
\end{eqnarray}
Using Theorem ~\ref{capa}, we have
\begin{eqnarray}
C^{(p)}(P_{X^{(n)}+1},P_{X^{(n)}})
&=&\sum_{\nu=1}^{\infty}\left(\frac{2}{n+1}\right)^{2\nu}\frac{1}{2\nu(2\nu-1)}\sum_{i=0}^{n+1}\left(i-\frac{n+1}{2}\right)^{2\nu}P_{X^{(n+1)}}(i)\nonumber\\
&& \qquad -[\log (2) -H(p)] \\
&=&\sum_{i=0}^{n+1}\sum_{\nu=1}^{\infty}\frac{2^{2\nu}}{2\nu(2\nu-1)} \left(\frac{i}{n+1}-\frac{1}{2} \right)^{2\nu}P_{X^{(n+1)}}(i) \nonumber\\
&& \qquad -[\log (2) -H(p)] \\
&\stackrel{a}{=}&\sum_{i=0}^{n+1} \left[ \log (2) -H\left(\frac{i}{n+1}\right) \right] P_{X^{(n+1)}}(i)
+H(p)-\log (2) ~~~~~~~~~~ \\
&=&H(p)-\sum_{i=0}^{n+1}H\left(\frac{i}{n+1}\right)P_{X^{(n+1)}}(i) \\
&\stackrel{b}{=}&
\sum_{i=0}^{n+1}\hat{H} \left(\frac{i}{n+1} \right)P_{X^{(n+1)}}(i),
\end{eqnarray}
where `$a$' follows by using \eqref{bin1} and `$b$' follows by using \eqref{hhat}.
To prove the lower bound, we have
\begin{eqnarray}
C^{(p)}(P_{X^{(n)}},P_{X^{(n)}+1})
&\stackrel{a}{\geq} &
\sum_{i=0}^{n+1}\sum_{k=1}^{2l+1}F^{(k)}(p){ \left(\frac{i}{n+1}-p \right)}^k P_{X^{(n+1)}}(i)  \\
&=&\sum_{k=1}^{2l+1}F^{(k)}(p)\sum_{i=0}^{n+1}{\left(\frac{i}{n+1}-p\right)}^kP_{X^{(n+1)}}(i)\\
\label{dummy3}
&=&\sum_{k=1}^{2l+1}F^{(k)}(p)(n+1)^{-k} \mu_k^{(n+1)},
\end{eqnarray}
where `$a$' holds for all nonnegative integers $l$ using \eqref{bound}.
\end{proof}

The following lemma shows that unlike the continuous case, EPI may not always hold.

\begin{lemma}
For $p \neq 0.5$, EPI does not hold for all $n$.
\end{lemma}
\begin{proof}
It suffices to show that  
\begin{equation}
\myexp^{2H[B(2,p)]}\leq \myexp^{2H[B(1,p)]}+\myexp^{2H[B(1,p)]} ~~~ \forall ~~ p,
\end{equation}
with equality if and only if $p=0.5$.
In other words, we need to show that
\begin{equation}
H[B(2,p)]-H[B(1,p)]-\frac{1}{2}\log (2) <0 ~~~ \forall ~~ p\neq 0.5.
\end{equation}
Using Lemma \ref{JSD} and \eqref{bin1}, we have
\begin{eqnarray}
H[B(2,p)]-H[B(1,p)]&=&H(p)-2p(1-p)\log (2),\\
H(p)&\leq & \log (2) -2 \left(p- 0.5 \right)^2.
\end{eqnarray}
Therefore,
\begin{eqnarray}
H[B(2,p)]-H[B(1,p)]-\frac{\log(2)}{2} &\leq& \frac{\log(2)}{2} -2\left(p-0.5\right)^2
-2p(1-p)\log (2) ~~~~~ \\
&=& 2\left(p-0.5\right)^2 \left[ \log (2)-1 \right] \\
&<& 0  \mbox{ if $p\neq 0.5 $. }
\end{eqnarray}
In other words, EPI holds for Binomial distributions $B(n,p)$ for all $n$ only if $p=0.5$.
\end{proof}

\begin{figure}
\centering
\includegraphics[height=3.5in]{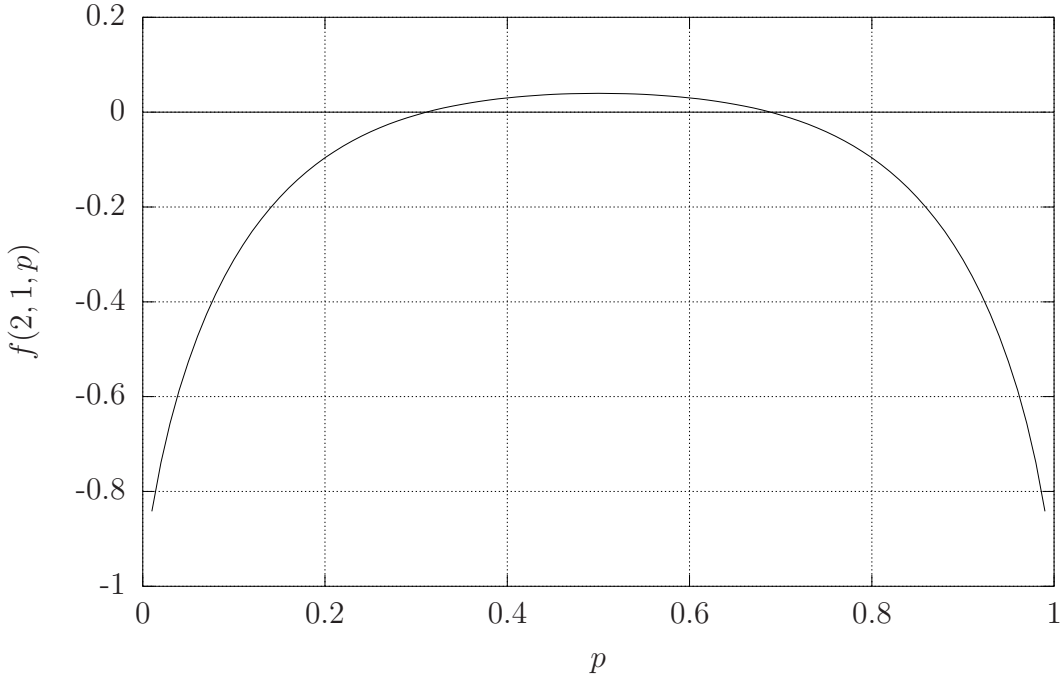}
\caption{ Plot of $\myexp^{2H[B(3,p)]}-\left\{\myexp^{2H[B(2,p)]}+\myexp^{2H[B(1,p)]}\right\}$
versus $p$.}
\label{fig}
\end{figure}

For the case $m=1$ and $n=2$, Fig. \ref{fig} shows the plot of
\beq
f(m,n,p)\triangleq \myexp^{2H[B(m+n,p)]}-\left\{\myexp^{2H[B(m,p)]}+\myexp^{2H[B(n,p)]}\right\}
\enq
as a function of $p$.
Note that EPI is satisfied for $p$ close to $0.5$, while EPI does not hold if $p$ is close to $0$ or $1$. 

This leads us to the question that for a given $p$, what should $m,n$ be such that the
EPI would hold. The main theorem of this section answers this question.
\begin{theorem}
\label{thm3}
\begin{equation}
H[B(n+1,p)]-H[B(n,p)]\geq \frac{1}{2}\log \left( \frac{n+1}{n} \right)
~~~ \forall ~~  n\geq n_0(p).
\end{equation}
Several candidates of $n_0(p)$ are possible such as $n_0(p) = 4.44 ~ \omega(p) + 7$
and $n_0(p) = \omega(p)^2 + 2.34 ~ \omega(p) + 7$.
\end{theorem}
\begin{proof}
See Appendix \ref{append1}.
\end{proof}

\subsection{Lower bound for the entropy of the Binomial distribution}

Unlike the asymptotic expansion of $H[B(n,p)]$ given in Ref. \cite{Knessl},
we give non-asymptotic lower bound to it. Let
\begin{equation}
\Gamma_{l}(j)\triangleq\sum_{k=1}^{2l+1}F^{(k)}(p)j^{-k} \mu_k^{(j)}.
\end{equation}
We have
\beq
H(X^{(j)})-H(X^{(j-1)}) \geq \Gamma_{l}(j),
\enq
where $X^{(n)}=B(n,p)$.

Using Fa\`{a} di Bruno's formula \cite{faadi}, we have
\begin{equation}
\mu_k^{(j)}=\sum \frac{k!}{i_1!(g_1!)^{i_1} \cdots i_s!(g_s!)^{i_s}} \kappa_{g_1}^{i_1} \cdots \kappa_{g_s}^{i_s}  j^{i_1+i_2+ \cdots + i_s},
\end{equation}
where $\kappa_{g}$ is the $g$-th cumulant of the Bernoulli distribution and
\beq
i_1g_1+i_2 g_2+\cdots+i_s g_s=k.
\enq
The summation is over all such partitions of $k$.
We have
\begin{eqnarray}
\Gamma_{l}(j) &=& \sum_{k=1}^{2l+1}F^{(k)}(p)j^{-k} \mu_k^{(j)} \\
      	&=&\sum_{w=1}^{2l}j^{-w}c(w),
\end{eqnarray}
where
\begin{equation}
c(w)=\sum \frac{k!}{i_1!(g_1!)^{i_1} \cdots i_s!(g_s!)^{i_s}} \kappa_{g_1}^{i_1} \cdots \kappa_{g_s}^{i_s} F^{(k)}(p)
\end{equation}
where the summation is over all such $(i_1,i_2,\cdots,i_s ; ~g_1,g_2,\cdots,g_s)$ such that
\beq
i_1(g_1-1)+i_2(g_2-1)+\cdots+i_s(g_s-1)=w.
\enq

Now
\begin{eqnarray}
H(X^{(n)})&=&H(X^{(0)})+\sum_{j=1}^{n}\left[H(X^{(j)})-H(X^{(j-1)})\right]\\
&\geq& \sum_{j=1}^{n} \Gamma_l(j), ~~~~~ \mbox{ since $H(X^{(0)}) = 0$} \\
&= & \sum_{j=1}^{n}\sum_{w=1}^{2l}j^{-w}c(w)\\
&= & \sum_{w=1}^{2l}c(w) \sum_{j=1}^{n} j^{-w}.
\end{eqnarray}
Note that $\sum_{j=1}^{n} j^{-w}$ are the {\it Generalized Harmonic Numbers} (see
for example Ref. \cite{conway}).

As an example, we compute the lower bound for $H(X^{(n)})$ for $l=1$.
We have
$c(1) = \kappa_{2}F^{(2)}(p)$ and $c(2) = 3\kappa_{2}^{2}F^{(4)}(p)+\kappa_{3}F^{(3)}(p)$.
The first and second cumulants of Bernoulli distribution is given by
$\kappa_2 = p(1-p)$ and $\kappa_3 = p(1-p)(1-2p)$.
This gives
$c(1) = 1/2$ and $c(2) = [1-p(1-p)]/[12p(1-p)]$ and we get the lower bound as
\begin{eqnarray}
H(X^{(n)})
\geq  \frac{1}{2} \left(1 + {1 \over 2} + \cdots {1 \over n} \right) +
\frac{1-p(1-p)}{12p(1-p)} \left( 1 + {1 \over 2^2} + \cdots {1 \over n^2} \right).
\end{eqnarray}

\section{EPI for the sum of IID}

We showed in the previous section that EPI holds for the pair $(B(n,p),B(m,p))$
for all $m,n \geq n_0(p)$.
The question naturally arises whether EPI holds for all such discrete random variables that
can be expressed as sum of IID random variables. 
Let $X^{(n)}$ be a discrete random variable such that
\beq
X^{(n)} \triangleq X_1+X_2+\cdots+X_{n},
\enq
where $X_i$'s are IID random variables and $\sigma^2$ is the variance of $X_1$.
We shall use the asymptotic expansion due to Knessl  \cite{Knessl}.
\begin{lemma}[Knessl \cite{Knessl}]
\label{lemma-knessl}
For a random variable $X^{(n)}$, as defined above, having finite moments, we have as $n \to \infty$,
\begin{equation}
\label{kness}
g(n) \triangleq
H(X^{(n)}) - \frac{1}{2} \log(2 \pi \myexp n\sigma^2 ) \sim -\frac{\kappa_3^2}{12 \sigma^6} \frac{1}{n}+ \sum_{l=1}^{\infty} \frac{\beta_l}{n^{l+1}},
\end{equation}
where $\kappa_j$ is the $j$th cumulant of of $X_1$.
If  $\kappa_3 = \kappa_4 = \cdots = \kappa_N = 0$ but $\kappa_{N+1} \neq 0$, then
\begin{equation}
g(n) \sim - \frac{\kappa_{N+1}^2}{2(N+1)!\sigma^{2N+2}}n^{1-N}+
\sum_{l=N-1}^{\infty} \frac{\beta_l}{n^{l+1}}.
\end{equation}
\end{lemma}
Note that the leading term in the asymptotic expansion is always negative.
We also note using Lemma \ref{lemma-knessl} that as $n \to \infty$,
\begin{equation}
\label{boundentropy}
H(X^{(n)})< \frac{1}{2}\log (2\pi \myexp n\sigma^2).
\end{equation}
To see this, we invoke the definition of the asymptotic series to get
\beq
g(n) =  - \frac{\kappa_{N+1}^2}{2(N+1)!\sigma^{2N+2}}n^{1-N}
+ {\beta_{N-1} \over n^N} + o\left( {1 \over n^N} \right).
\enq
From the definition of the ``little-oh" notation, we know that given any $\epsilon > 0$, there exists a
$L(\epsilon)>0$ such that for all $n> L(\epsilon)$,
\beq
g(n) =  - \frac{\kappa_{N+1}^2}{2(N+1)!\sigma^{2N+2}}n^{1-N}
+ {\beta_{N-1} + \epsilon \over n^N}.
\enq
Choosing $n$ large enough, we get the desired result.

\subsection{Asymptotic case}

We first consider the case of the pair $(X^{(n)},X^{(m)})$ when both $m,n$ are large and have
the following result.

\begin{theorem}
\label{asym}
There exists a $n_0\in \mathbb{N}$ such that
\begin{equation}
\label{iin}
\myexp^{2H(X^{(m)}+X^{(n)})} \geq \myexp^{2H(X^{(m)})}+\myexp^{2H(X^{(n)})}
\end{equation}
for all $m,n\geq n_0$.
\end{theorem}
\begin{proof}
We shall prove the sufficient condition for the EPI to hold (as per Lemma \ref{superadd}) and show that
\begin{equation}
\label{hxn}
H(X^{(n+1)})-H(X^{(n)}) \geq \frac{1}{2} \log \left( \frac{n+1}{n} \right)
\end{equation}
for $n\geq n_0$ for some $n_0\in \mathbb{N}$.

Let us take the first three terms in the above asymptotic series as
\begin{equation}
\label{threeterm}
g(n) \sim -{C_1 \over n^{k_1}} + {C_2 \over n^{k_2}} + {C_3 \over n^{k_3}}
\end{equation}
where $0<k_1< k_2< k_3$ and $C_1$ is some non-zero positive constant, and hence,
\begin{equation}
g(n) + {C_1 \over n^{k_1}} - {C_2 \over n^{k_2}} - {C_3 \over n^{k_3}} =
o\left( {1 \over n^{k_3}} \right).
\end{equation}
and given any $\epsilon > 0$, there exists a
$L(\epsilon)>0$ such that for all $n> L(\epsilon)$,
\begin{equation}
\left| g(n) + {C_1 \over n^{k_1}} - {C_2 \over n^{k_2}} - {C_3 \over n^{k_3}} \right|
\leq \epsilon \left|{1 \over n^{k_3}} \right|.
\end{equation}

Therefore, we have the following inequality
\begin{equation}
\label{ineqsn}
\frac{-C_1}{n^{k_1}}+\frac{C_2}{n^{k_2}}+\frac{C_3-\epsilon}{n^{k_3}} \leq g(n) \leq \frac{-C_1}{n^{k_1}}+\frac{C_2}{n^{k_2}}+\frac{C_3+\epsilon}{n^{k_3}}.
\end{equation}

From inequality \eqref{ineqsn}, we can say by using the lower and upper bounds respectively for $g(n+1)$ and $g(n)$ that,
\begin{align}
g(n+1)-g(n) \geq C_1\left[\frac{1}{n^{k_1}} - \frac{1}{(n+1)^{k_1}}\right]  &+ C_2\left[\frac{1}{(n+1)^{k_2}} - \frac{1}{n^{k_2}}\right] \nonumber \\
&+ \left[\frac{C_3-\epsilon}{(n+1)^{k_3}}- \frac{C_3 + \epsilon}{n^{k_3}}\right].
\end{align}
From the above expression, we can clearly see that the first term is strictly positive and is
$O\left(1/n^{k_1+1}\right)$. The second and third terms (their signs are irrelevant) are of the order
$O(1/n^{k_2+1})$ and $O(1/n^{k_3})$ respectively.
It is clear that there exists some positive integer $n_0$ such that for all $n\geq n_0$,
first (positive) term will dominate and the other two terms will be negligible compared to the first and hence $g(n+1)-g(n)\geq 0$.
\end{proof}

\subsection{Semi-asymptotic case}

We now consider the pair $(X^{(n)},X^{(m)})$ where $n \to \infty$ and $m$ is fixed.
We already know from the previous result that the EPI holds when both $m$ and $n$ are large.

We start by writing an asymptotic expansion of
\begin{equation}
f(m,n) \triangleq \myexp^{2H(X^{(m+n)})} - \myexp^{2H(X^{(n)})} - \myexp^{2H(X^{(m)})},
\end{equation}
by using Knessl's result in Lemma \ref{lemma-knessl}
in which an asymptotic expansion for the entropy of $X^{(n)}$ is derived as
\begin{equation}
H(X^{(n)}) \sim \frac{1}{2} \log(2 \pi \myexp n\sigma^2) + \sum_{l=0}^{\infty} \frac{\beta_l}{n^{l+1}}.
\end{equation}
Let
\begin{equation}
g(n) \triangleq \sum_{l=0}^{\infty} \frac{\beta_l}{n^{l+1}}.
\end{equation}
We have
\begin{eqnarray}
\myexp^{2H(X^{(n)})} &=& (2\pi \myexp n\sigma^2) \myexp^{g(n)},\\
\myexp^{2H(X^{(m+n)})} &=& [2\pi \myexp (m+n)\sigma^2] \myexp^{g(m+n)},
\end{eqnarray}
and we can rewrite
\begin{align}
\label{dummy4}
f(m,n) = \left[2\pi \myexp m\sigma^2 - \myexp^{2H(X^{(m)})}\right] + 2\pi &
\myexp (m+n)\sigma^2\left[ \myexp^{g(m+n)}-1\right] \nonumber \\
&- 2\pi \myexp n\sigma^2\left[ \myexp^{g(n)}-1\right].
\end{align}
The first term in the above equation is a constant since it depends only
on $m$ and the second term can be written as
\begin{equation}
2\pi \myexp (m+n)\sigma^2\left\{ \myexp^{\left[\frac{\beta_0}{m+n} + \frac{\beta_1}{(m+n)^2} + o\left(\frac{1}{(m+n)^2}\right)\right]} -1 \right\},
\end{equation}
which can be expanded into
\begin{equation}
2\pi \myexp (m+n)\sigma^2\left\{ \frac{\beta_0}{m+n} + \frac{2\beta_1 + \beta_0^2}{2(m+n)^2} + o\left[\frac{1}{(m+n)^2}\right] \right\}.
\end{equation}
Similarly, the third term can be written as
\begin{equation}
-2\pi \myexp n\sigma^2\left[ \frac{\beta_0}{n} + \frac{2\beta_1 + \beta_0^2}{2n^2} + o\left(\frac{1}{n^2}\right) \right].
\end{equation}
Using the above two expressions,
$o\left[1/(m+n)^2\right]=o\left(1/n^2\right)$, and \eqref{dummy4}, we get
\begin{equation}
f(m,n) = [2\pi \myexp m\sigma^2 - \myexp^{2H(X^{(m)})}]+2\pi \myexp \sigma^2 \left[ \frac{2\beta_1 + \beta_0^2}{2} \left( \frac{1}{m+n} -\frac{1}{n} \right) + o\left( \frac{1}{n} \right) \right],
\end{equation}
where the terms
\begin{equation*}
2\pi \myexp \sigma^2 \left[o\left( \frac{1}{n} \right)\right] \mbox{ and } 2\pi \myexp \sigma^2 \left( \frac{2\beta_1 + \beta_0^2}{2}\right) \left( \frac{1}{m+n} -\frac{1}{n} \right)
\end{equation*}
can be made arbitrarily small as $n \to \infty$.
Therefore, for large enough $n$, we can see that the first term dominates over other terms and moreover, $f(n,m)\geq 0$ if $2\pi \myexp m\sigma^2 - \myexp^{2H(X^{(m)})} > 0$.
Therefore,
\begin{equation}
\myexp^{2H(X^{(m)}+X^{(n)})} \geq \myexp^{2H(X^{(m)})}+\myexp^{2H(X^{(n)})}
\end{equation}
if $n\to\infty$ and
\begin{equation}
H(X^{(m)})< \frac{1}{2}\log [2\pi \myexp m\sigma^2].
\end{equation}
It follows from \eqref{boundentropy} that the above inequality holds for sufficiently large $m$.
For the Binomial distribution $B(m,p)$, EPI will hold for all such $p$ that satisfy
\begin{equation}
H[B(m,p)] < \frac{1}{2}\log [2\pi \myexp mp(1-p)].
\end{equation}
The above relation is not true for all  $p$ and $m$.

\section{Discussion and Conclusions}

We show that how our results can be used to improve a bound by Tulino and Verd\'{u}
\cite{tulino} under special cases.

\subsection{Improvement on a bound by Tulino and Verd\'{u}}

Let $X_i$, $i=1,2,...,n$ be discrete IID random variables and $Z_i$, $i=1,2,...,n$, be IID random
variables as well with $Z_1 \sim\mathcal{N}(0,\sigma^2)$.
Let
\beqa
S^{(n)} & = & \sum_{i=1}^{n}(X_i+Z_i) \\
X^{(n)} & = & \sum_{i=1}^{n} X_i.
\enqa

Let
\beq
\label{eq-D}
D(Y) = 0.5 \log(2\pi \myexp \sigma_Y^2)-h(Y),
\enq
where $Y$ is a random variable with density and variance $\sigma_Y^2$.
Tulino and Verd\'{u}  \cite{tulino} interpreted $D(Y)$ as the non-Gaussianess of the random variable
$X$ and showed that the non-Gaussianess increases by having more random variables, i.e.,
\beq
D(S^{(n)}) \leq D(S^{(n-1)}).
\enq
Expanding it using \eqref{eq-D}, we get
\begin{equation}
\label{tulino-bound}
h(S^{(n)})-h(S^{(n-1)}) \geq \frac{1}{2}\log \left(\frac{n}{n-1} \right).
\end{equation}

We show that for sufficiently large $n$, this bound can be made tighter for small $\sigma$, i.e.,
\begin{equation}
\lim_{\sigma\to 0} ~ \left[h(S^{(n)})-h(S^{(n-1)}) \right] \geq \log \left(\frac{n}{n-1}\right).
\end{equation}

Let
\beq
I_n \triangleq I(X^{(n)} ; S^{(n)}) = h(S^{(n)})- 0.5 \log(2\pi \myexp n\sigma^2).
\enq
Note that using Lemma 6 in Ref. \cite{guo-2005}
\begin{equation} H(X^{(n)})=\lim_{\sigma\to 0}(I_n).
\end{equation}
We know that for sufficiently large $n$, the EPI holds and
\beq
H(X^{(n)})-H(X^{(n-1)})\geq\frac{1}{2}\log  \left(\frac{n}{n-1} \right).
\enq
Therefore,  
\begin{equation}
\lim_{\sigma\to 0} ~ (I_n-I_{n-1})\geq\frac{1}{2}\log \left(\frac{n}{n-1} \right)
\end{equation}
for sufficiently large $n$. The above limit is due to  \cite{guo-2005}.
On the other hand,
\beq
I_n-I_{n-1}=h(S^{(n)})-h(S^{(n-1)})-\frac{1}{2}\log\left(\frac{n}{n-1}\right).
\enq
Hence,  
\begin{equation}
\label{our-bound}
\lim_{\sigma\to 0} ~ [h(S^{(n)})-h(S^{(n-1)})]\geq \log \left(\frac{n}{n-1} \right)
\end{equation}
for sufficiently large $n$. Comparing \eqref{our-bound}
with \eqref{tulino-bound}, we note that our bound is tighter by a factor of $2$.

\subsection{Conclusions}

In conclusion, we have expanded the set of pairs of Binomial distributions for which the
EPI holds. We identified a threshold that is a function of the probability of success beyond
which the EPI holds. We further show that EPI would hold for discrete random variables that 
can be written as sum of IID random variables.

It would be interesting to know if $C^{(p)}(P_{X^{(n)}+1},P_{X^{(n)}})$ for $X^{(n)} = B(n,p)$
is a concave function in $p$. It would also be of interest to know that for a given $p \in (0,0.5)$
if $H[B(n+1,p)]-H[B(n,p)] - 0.5 \log ( 1+1/n )$ would have a single zero
crossing as a function of $n$ when $n$ increases from $1$ to $\infty$.

\newpage

\bibliographystyle{IEEETran}
\bibliography{ref}

\newpage

\appendix

\section{Proof of Theorem \ref{thm3}}
\label{append1}
We prove that
\begin{equation}
\label{dummy1}
H[B(n+1,p)]-H[B(n,p)]
\geq \frac{1}{2}\log \left( \frac{n+1}{n} \right) ~~ \forall ~ n\geq n_0(p).
\end{equation}
Using \eqref{dummy3}, we have
\begin{eqnarray}
H[B(n+1,p)]-H[B(n,p)]&=&
C^{(p)}(P_{X^{(n)}},P_{X^{(n)}+1})\\
& \geq & \sum_{k=1}^{2l+1}F^{(k)}(p)(n+1)^{-k} \mu_{k}^{(n+1)}.
\end{eqnarray}
Let
\beq
r \triangleq p-1/2.
\enq
We have the first seven central moments of $B(n,p)$ as  
\begin{eqnarray}
\label{cenmoments}
\mu_2^{(n)} &=&\frac{1}{4}n (1 - 4r^2),\\
\mu_3^{(n)} &=&-\frac{1}{2}nr (1- 4 r^2),\\
\mu_4^{(n)} &=&\frac{1}{16} n (1 - 4 r^2) [-2 + 24 r^2 + 3 n (1 - 4 r^2)],\\
\mu_5^{(n)} &=&-\frac{1}{4}n r (1- 4 r^2)[-4 +24 r^2 + 5 n (1 -4 r^2)],\\
\mu_6^{(n)} &=&\frac{1}{32} n (1 - 4 r^2) [15 n^2 (1 - 4 r^2)^2 +
   16 (1 - 30 r^2 + 120 r^4) \nonumber \\
   & & ~~~~~ - 10 n (3 - 64 r^2 + 208 r^4)],\\
\mu_7^{(n)}  &=&-\frac{1}{32}n r (1-4 r^2) [105 n^2 (1 - 4 r^2)^2
-14 n (17 - 200 r^2 + 528 r^4), \\
&& ~~~~~ + 8 (17 - 240 r^2 + 720 r^4)].
\end{eqnarray}

Let $t=\omega(p)=16r^2/(1-4r^2)$ and hence, $r^2=t/[4(t+4)]$.
Note that $r^2$ $\in$ $[0,1/4)$ and $t$ $\in$ $[0,\infty)$.
The above seven central moments contain only
even powers of $r$ and hence, can be written as a function of $t$.

We upper bound the right hand side of \eqref{dummy1} as
\beq
\log \left( {n+1 \over n} \right) \leq {1 \over n} - {1 \over 2 n^2} + {1 \over 3 n^3}.
\enq

Define
\begin{eqnarray}
f(n,t)\triangleq\sum_{k=1}^{7}F^{(k)}\left[ \sqrt{t \over 4(t+4)} + {1 \over 2} \right] (n+1)^{-k} \mu_k^{(n+1)}
- {1 \over n} + {1 \over 2 n^2} - {1 \over 3 n^3}.
\end{eqnarray}
Proving \eqref{dummy1} is equivalent to showing that $f(n,t) \geq 0$ $\forall$ $n > n_0(p)$.
Simplifying
\beqa
f(n,t) & = & {1 \over 420 (n+1)^6 n^3} \Big[
35 n^7 t+(315 t+35 t^2+70) n^6+ \nonumber \\
& & ~~~~ (-2989 t-721 t^3-3339 t^2-315)n^5 + \nonumber \\
& & ~~~~ (721 t-546 t^4-826+371 t^2-1568 t^3) n^4+ \nonumber \\
& & ~~~~ (-135 t^2-157 t^3-10 t^5-826-90 t-66 t^4) n^3- \nonumber \\
& & ~~~~ 630 n^2-315 n - 70 \Big].
\enqa

Define
\beq
g(n,t) \triangleq 420 (n+1)^6 n^3 f(n,t).
\enq
A simple but elaborate calculation yields
$g(4.44t+7+m,t) \approx
35t m^7 + (1122.8 t^2+2030 t+70) m^6 +
(14700.90 t^3 + 52210.20 t^2 + 48120.80 t + 2625) m^5+
(1.01 t^4 + 5.32 t^3 + 9.57 t^2 + 6.06 t + 0.40) 10^5 m^4+
(3.85 t^5 + 26.94 t^4 + 72.32 t^3 + 88.61 t^2+ 43.61 t  + 3.02) 10^5 m^3+
(7.76 t^6+68.23 t^5+247.042 t^4+456.97 t^3+433.17 t^2+176.77 t+11.80) 10^5 m^2+
(6.47 t^7+70.91 t^6+338.88 t^5+880.98 t^4+1297.85 t^3+
1030.51 t^2+361.59 t+20.14) 10^5 m + (0.15 t^8+56.29 t^7+709.80 t^6+3485.03 t^5+8728.40 
t^4+11955.74 t^3+8613.06 t^2+2628.77 t+64.15)10^4 $.

Note that all the coefficients are positive and hence, $f(4.44t+7+m,t) \geq 0$ for all $m \geq 0$
or $f(n,t) \geq 0$ for all $n \geq 4.44 t + 7$.
A more careful choice would yield all coefficients to be positive for $n \geq 4.438 t + 7$.
Yet another choice that would yield all coefficients as positive would be
$n \geq t^2+2.34 t+7$. Note that this choice would be better for $0 < t < 2.1$ and, in particular,
for $t=1$, the first choice yields (after constraining $n$ to be a natural number)
$n \geq 12$ while the second one yields $n \geq 11$.

Further refinements are also possible. For example, the expansion of
$f(7+m, t/[4(1+t)])$ yields positive coefficients again. Such a choice constrains
$\omega(p)$ $\in$ $(0,1/4)$ and we get $n_0(p) = 7$.

\end{document}